%% file: ANA-HIGP-2024-52-PAPER.tex
\documentclass[PAPER, atlasdraft=false, UKenglish, cernpreprint, texmf, orcidlogo]{atlasdoc}
\usepackage{atlaspackage}
\usepackage{multirow}
\usepackage{atlasbiblatex}

\usepackage{atlasphysics}
\graphicspath{{logos/}{figures/}}

\usepackage{ANA-HIGP-2024-52-PAPER-defs}

\usepackage{rotating}

\AtlasTitle{Study of $t\bar{t}H$ and $tH$ production in the $H\to\tau\tau$ channel in $pp$ collisions at $\sqrt{s}=13$~TeV and 13.6~TeV with the ATLAS detector}

\AtlasAbstract{%
A study of the production of the Higgs boson in association with either a top-quark pair ($t\bar{t}H$) or a single top quark ($tH$) in the $\tau$-lepton-pair decay channel is presented.
The analysis relies on final states featuring fully hadronic decays of the top quarks and the $\tau$-leptons.
It employs data samples of proton--proton collisions at
$\sqrt{s}=13$ and 13.6~TeV recorded with the ATLAS detector at the CERN Large
Hadron Collider and corresponding to integrated luminosities of 140~fb$^{-1}$ and 161~fb$^{-1}$, respectively.
The measured signal strength ($\mu$), defined as the measured cross-section normalised to the Standard Model prediction, is
$\mu_{t\bar{t}H}=1.51^{+0.71}_{-0.62}$ for $t\bar{t}H$ and $\mu_{tH}=-0.4^{+5.6}_{-5.1}$ for $tH$.
Additionally,  the $t\bar{t}H$ cross-section is measured differentially in three bins of the Higgs boson transverse momentum in the
simplified template cross-section framework.
}

\AtlasRefCode{HIGP-2024-52}

\PreprintIdNumber{CERN-EP-2026-229}

\AtlasJournal{Phys.\ Lett.\ B.}
\AtlasCoverEgroupAnalysisTeam{atlas-higp-2024-52-analysis-team@cern.ch}

\hypersetup{pdftitle={ATLAS document},pdfauthor={The ATLAS Collaboration}}

\begin{document}

\maketitle

\tableofcontents

\section{Introduction}
\label{sec:intro}
%

Since the discovery of the Higgs boson with a mass of approximately 125~\GeV by the ATLAS~\cite{HIGG-2012-27} and CMS~\cite{CMS-HIG-12-028} Collaborations in 2012,
the measurement of its properties has been one of the primary goals of the physics programme at the Large Hadron Collider
(LHC)~\cite{Evans:2008zzb}. So far no significant deviation from the Standard Model (SM) has been observed~\cite{HIGP-2025-16,CMS-HIG-21-018}.

The Yukawa coupling of the Higgs boson with fermions was established with the study of the decay into a $\tau^+\tau^-$ pair\footnote{For simplicity, a $\tau^+\tau^-$~pair is denoted by $\tau\tau$~throughout this paper.} ($\Htt$), the decay mode with the largest branching ratio of all leptonic Higgs boson decays.
Evidence for \Htt was reached using the LHC Run 1 data independently by ATLAS~\cite{HIGG-2013-32} and CMS~\cite{CMS-HIG-13-004},  and their combination resulted in the observation of this decay channel~\cite{HIGG-2015-07}.
Current results reach a precision of approximately 12\% in the measurement of the $\Htt$ production~\cite{HIGG-2022-07,CMS-HIG-19-010}.

The top-quark Yukawa coupling is of particular interest.  Its large mass of approximately 173~\GeV~\cite{Zyla:2020zbs}
implies a strong coupling with the Higgs boson,  suggesting that the top quark could play a special role in the electroweak symmetry breaking.
The measurement of the inclusive cross-section of the associated production of the Higgs boson with a top-quark pair (\ttH)
allows a  direct determination of the top-quark Yukawa coupling,
relying on fewer theoretical assumptions than inferring the value of the coupling from its contributions to the loop-induced Higgs boson gluon--gluon fusion production or decay into pairs of photons.
The \ttH process was observed with a partial Run 2 data sample by both ATLAS~\cite{HIGG-2018-13} and
CMS~\cite{CMS-HIG-17-035} using a combination of different channels, and current measurements of the \ttH production cross-section have a precision of approximately 20\%~\cite{HIGP-2025-16,CMS-HIG-21-018}.
In the simplified template cross-section (STXS)~\cite{Andersen:2016qtm,Berger:2019wnu} formalism,   \ttH production is measured differentially in bins of the Higgs boson transverse momentum (\pTH), a variable sensitive to possible new physics effects, such as $CP$-mixing in the top-quark Yukawa coupling or an anomalous Higgs boson self-coupling~\cite{Houches2019,Berger:2019wnu}.

Higgs boson production in association with a single top quark ($tH$) at the LHC occurs primarily via $tHqb$ ($t$-channel) and $tWH$.
The $s$-channel $tHb$ production has a very low cross-section and is neglected in this paper.
The \tH process has been searched for in the $H \to \gamma\gamma$~\cite{HIGP-2024-24,CMS-HIG-19-015},
$H \to b\bar{b}$~\cite{HIGP-2024-03,CMS-HIG-19-011}
and multilepton (with several light or $\tau$-leptons originating from both the decays of the Higgs boson and the top quarks) channels~\cite{HIGP-2026-05,HIGP-2024-08,HIGP-2024-03,CMS-HIG-19-008}.
Current results on $tH$ production have uncertainties at the level of 3-4 times the SM prediction,  with the measurements being compatible with,  but slightly above the expectations.

This paper presents a simultaneous measurement of the \ttH and \tH production relying on the \Htt decay with fully hadronic final states,
using the complete LHC Run 2 data sample together with the Run 3 data acquired during 2022 to 2024.
While the inclusive $\ttH$ cross-section is primarily sensitive to the absolute strength of the top-quark Yukawa coupling,  the $tH$ process is affected by the relative phase between the Higgs couplings to the top quark and to the $W$ boson.  Therefore,  a simultaneous study  of $\ttH$  and $tH$ provide sensitivity to potential \(CP\) violation effects in the interaction between the top quark and the Higgs boson.
Furthermore, exploring  \ttH production with \Htt with both top quarks and $\tau$-leptons decaying hadronically benefits from having an unambiguous Higgs boson candidate and very low contamination from other Higgs boson decays.
A measurement of $\ttH$ in a similar final state was performed in Ref.~\cite{HIGG-2022-07} including an STXS interpretation,
but this paper includes for the first time a \tH measurement in this channel.  %
In addition, the analysis benefits from the larger data sample and several analysis improvements.

The paper is structured as follows.
The ATLAS detector is presented in Section~\ref{sec:detector}.
The recorded data and the Monte Carlo (MC) simulation samples used in the analysis are described in Section~\ref{sec:dataMC}.
The definition of reconstructed objects is described in Section~\ref{sec:objects}.
The event selection is given in Section~\ref{sec:evtselection}.
The background estimation method and the systematic uncertainties are described in Sections~\ref{sec:bkg} and~\ref{sec:syst}, respectively.
The fit setup and results are presented in Section~\ref{sec:result}.


%
\section{ATLAS detector}
\label{sec:detector}
%

%
\newcommand{\AtlasCoordFootnote}{%
ATLAS uses a right-handed coordinate system with its origin at the nominal interaction point (IP)
in the centre of the detector and the \(z\)-axis along the beam pipe.
The \(x\)-axis points from the IP to the centre of the LHC ring,
and the \(y\)-axis points upwards.
Polar coordinates \((r,\phi)\) are used in the transverse plane,
\(\phi\) being the azimuthal angle around the \(z\)-axis.
The pseudorapidity is defined in terms of the polar angle \(\theta\) as \(\eta = -\ln \tan(\theta/2)\) and is equal to the rapidity
$ y = \frac{1}{2} \ln \left( \frac{E + p_z}{E - p_z} \right) $ in the relativistic limit.
Angular distance is measured in units of \(\Delta R \equiv \sqrt{(\Delta \eta)^{2} + (\Delta\phi)^{2}}\).}

The ATLAS experiment~\cite{PERF-2007-01,GENR-2019-02} at the LHC is a multipurpose particle detector
with a forward--backward symmetric cylindrical geometry and a near \(4\pi\) coverage in
solid angle.\footnote{\AtlasCoordFootnote}
It consists of an inner tracking detector surrounded by a thin superconducting solenoid
providing a \qty{2}{\tesla} axial magnetic field, electromagnetic and hadronic calorimeters, and a muon spectrometer.
The inner tracking detector covers the pseudorapidity range \(|\eta| < 2.5\).
It consists of silicon pixel, silicon microstrip, and transition radiation tracking detectors.
Lead/liquid-argon (LAr) sampling calorimeters provide electromagnetic (EM) energy measurements
with high granularity within the region \(|\eta|< 3.2\).
A steel/scintillator-tile hadronic calorimeter covers the central pseudorapidity range (\(|\eta| < 1.7\)).
The endcap and forward regions are instrumented with LAr calorimeters
for EM and hadronic energy measurements up to \(|\eta| = 4.9\).
The muon spectrometer surrounds the calorimeters and is based on
three large superconducting air-core toroidal magnets with eight coils each.
The field integral of the toroids ranges between \num{2.0} and \qty{6.0}{\tesla\metre}
across most of the detector.
The muon spectrometer includes a system of precision tracking chambers up to \(|\eta| = 2.7\) and fast detectors for triggering up to \(|\eta| = 2.4\).
The luminosity is measured mainly by the LUCID--2~\cite{LUCID2} detector which is located close to the beampipe.
A two-level trigger system was used to select events~\cite{TRIG-2016-01,TRIG-2022-01}.
The first-level trigger is implemented in hardware and used a subset of the detector information
to accept events at a rate close to \qty{100}{\kHz}.
This is followed by a software-based trigger that
reduced the accepted rate of complete events to \qty{1.25}{\kHz} and \qty{3}{\kHz} on average in Run~2 and Run~3, respectively,
depending on the data-taking conditions.
A software suite~\cite{SOFT-2022-02} is used in data simulation, in the reconstruction
and analysis of real and simulated data, in detector operations, and in the trigger and data acquisition
systems of the experiment.


%
\section{Data and simulated event samples}
\label{sec:dataMC}
%

This analysis uses the data collected by the ATLAS detector in $pp$ collisions at the LHC during 2015--2018 (Run~2) at $\sqrt{s} = \SI{13}{\TeV}$,  corresponding to a total integrated luminosity of 140\,\ifb~\cite{DAPR-2021-01},  and during 2022--2024 (Run~3) at $\sqrt{s} = \SI{13.6}{\TeV}$,  corresponding to a total integrated luminosity of 161\,\ifb~\cite{ ATL-DAPR-PUB-2025-001}.
The uncertainty on the luminosity measurements is 0.83\% for Run~2 and 1.9\% for Run~3 obtained using the LUCID-2 detector for the primary luminosity measurements with complementary measurements from the inner detector and calorimeters.
Events were collected with triggers requiring the presence of two hadronically decaying $\tau$-leptons (\tauhad) and one additional jet in the event~\cite{TRIG-2016-01,TRIG-2022-01,TRIG-2019-02}.
The requirements imposed on the \pt of the $\tau$-leptons and jets in the final selection %
are tighter than those applied in the trigger-object reconstruction and trigger decision,  to be on the trigger efficiency plateau.

Simulated MC event samples are used to model the $\ttH$ and $\tH$ signals and to estimate the
backgrounds from SM processes.  A summary of the generators used for the simulation is shown in Table~\ref{tab:MCSamples}, and further details are provided below for the most relevant samples in the analysis.

\begin{table}[tbh]
\begin{center}
\caption{
The configurations used for event generation of signal and background processes, where $V$ denotes $W$ or $Z$ bosons.
The matrix element order refers to the order in the strong coupling constant of the perturbative calculation.
Tune refers to the underlying-event tune of the parton shower generator,  either A14~\cite{ATL-PHYS-PUB-2014-021} or the default in \SHERPA~\cite{Bothmann:2019yzt}.
All samples include leading-logarithm photon emission, either modelled by the parton shower generator or by \textsc{Photos}~\cite{Golonka:2005pn}.
The mass of the top quark ($m_t$) and SM Higgs boson were set to $\SI{172.5}{\GeV}$ and $\SI{125}{\GeV}$, respectively.
}
\label{tab:MCSamples}
\vspace{0.25cm}
\small
\setlength\tabcolsep{5pt}
\scalebox{0.9}{
\begin{tabular}{l l l l l l l}
\hline\hline
Physics process & Event generator & Matrix element order & Parton shower & PDF set & Tune  \\
\hline
\multirow{ 2}{*}{\ttH}                   & \multirow{ 2}{*}{\POWHEGBOX} & \multirow{ 2}{*}{NLO} & \multirow{ 2}{*}{\PYTHIA[8]} & \NNPDF[3.0nnlo] (Run 2) & \multirow{ 2}{*}{A14} \\
&  &  & & \PDFforLHC[21nlo] (Run 3) &  \\

$tH$  & \MGNLO & NLO & \PYTHIA[8] & \NNPDF[3.0nlo] & A14  \\ \hline
\Vjets & \SHERPA[2.2.14] & NLO &  \SHERPA &  \NNPDF[3.0nnlo] & \SHERPA  \\
$\ttbar$, single top          & \POWHEGBOX & NLO & \PYTHIA[8] & \NNPDF[3.0nlo] & A14  \\

$\ttbar Z/\gamma^\ast$       & \MGNLO & NLO & \PYTHIA[8]  &  \NNPDF[3.0nlo] &  A14 \\
$\ttbar W$             & \SHERPA[2.2.10] & NLO  & \SHERPA &  \NNPDF[3.0nnlo] & \SHERPA  \\
$t\bar t t\bar t$      &\MGNLO & NLO & \PYTHIA[8] & \NNPDF[3.1nlo] & A14 \\
$VV$, $qqVV$, $VVV$    & \SHERPA[2.2.2(1)]  & NLO & \SHERPA & \NNPDF[3.0nnlo] & \SHERPA   \\
\hline\hline
\end{tabular}
}
\end{center}
\end{table}

The production of \ttH events was modelled at next-to-leading order (NLO) in QCD using the
\POWHEGBOX[v2]~\cite{Hartanto:2015uka, Nason:2004rx,Frixione:2007vw,Alioli:2010xd,Hartanto:2015uka}
generator, which provided matrix elements at NLO in the strong coupling
constant \alphas in the five-flavour scheme.
As parton distribution function (PDF) set,  the \NNPDF[3.0nnlo]~\cite{Ball:2014uwa} (\PDFforLHC[21nlo]~\cite{Butterworth:2015oua}) was used to simulate the Run~2 (Run~3) data.
The events were interfaced to \PYTHIA[8]~\cite{Sjostrand:2014zea}
using the \NNPDF[2.3lo]~\cite{Ball:2012cx} PDF set.
The cross-section was calculated at NLO QCD and NLO electroweak (EW) accuracy using
\MGNLO as reported in Refs.~\cite{deFlorian:2016spz,karlberg2024adinterimrecommendationshiggs}.
The predicted value is
\(507^{+35}_{-50}\,\unit{\fb}\) at \(\rts = \qty{13}{\TeV}\) and \(570^{+44}_{-60}\,\unit{\fb}\) at \(\rts = \qty{13.6}{\TeV}\), where the uncertainties were estimated from
variations of PDF,  \alphas and the renormalisation and factorisation scales.

The production of \tH events is modelled using \MGNLO interfaced with \PYTHIA[8.309] using the \NNPDF[3.0nlo]  set for the matrix element.
The cross-section was calculated at NLO QCD accuracy using
\MGNLO as reported in Refs.~\cite{deFlorian:2016spz,karlberg2024adinterimrecommendationshiggs}.
The predicted values at $\sqrt{s}=13(13.6)$$~\SI{}{\TeV}$ are
$74.2^{+5.5}_{-11.4}$ $(83.6^{+6.3}_{-12.7})$~\SI{}{fb} for $t$-channel $qb\to tHq$ production,  and
$15.2 ^{+1.2}_{-1.4}$$(17.2 ^{+1.4}_{-1.6})$~\SI{}{fb} for $gb\to tHW$ production.

Other Higgs boson production modes (gluon--gluon fusion,  vector-boson fusion and in association with a vector boson) are simulated with \POWHEGBOX[v2] interfaced with \PYTHIA[8] for parton shower and non-perturbative effects and using \PDFforLHC[15nlo] (\PDFforLHC[21nlo]) for Run~2 (Run~3).
The predictions from the Monte Carlo samples are normalised using the most up-to-date theory predictions for the SM production cross-sections, as summarised in Ref. ~\cite{deFlorian:2016spz}.

Vector boson production in association with jets ($V$+jets) is simulated with the \SHERPA[2.2.14]~\cite{Bothmann:2019yzt} generator using NLO matrix elements for up to two partons, and leading order (LO) matrix elements for up to five partons calculated with the Comix~\cite{Gleisberg:2008fv} and \openloops~\cite{Buccioni:2019sur,Cascioli:2011va,Denner:2016kdg} libraries.
They are matched with the \sherpa parton shower~\cite{Schumann:2007mg} using the MEPS@NLO prescription~\cite{Hoeche:2011fd,Hoeche:2012yf,Catani:2001cc,Hoeche:2009rj} and the set of tuned parameters developed by the \sherpa authors.
The  \NNPDF[3.0nnlo] set of PDFs~\cite{Ball:2014uwa} is used and the samples are normalised to a next-to-next-to-leading order (NNLO) prediction~\cite{Anastasiou:2003ds}.

The production of \ttbar\ events is modelled at NLO in QCD using the \POWHEGBOX[v2]~\cite{Frixione:2007nw,Nason:2004rx,Frixione:2007vw,Alioli:2010xd} generator at NLO with the \NNPDF[3.0nlo]~\cite{Ball:2014uwa} PDF set and the \hdamp\ parameter\footnote{The \hdamp\ parameter is a resummation damping factor and one of the parameters that controls the matching of \powheg matrix elements to the parton shower and thus effectively regulates the high-$\pt$ radiation against which the \ttbar\ system recoils.} set to 1.5~$m_t$~\cite{ATL-PHYS-PUB-2016-020}.
The events are interfaced to \PYTHIA[8.230] to model the parton shower, hadronisation, and underlying event, and using the \NNPDF[2.3lo] set of PDFs~\cite{Ball:2012cx}.
Additional MC samples are used to model processes with smaller impact in the analysis (such as single top,  $\ttbar W$,  $\ttbar Z$, $\ttbar\ttbar$,  or multi-boson production), whose generator and settings are shown in Table~\ref{tab:MCSamples}.

All MC samples produced with \POWHEGBOX and \MGNLO used the \PYTHIA A14 underlying event tune~\cite{ATL-PHYS-PUB-2014-021},
and the decays of bottom and charm hadrons were performed by \EVTGEN~\cite{Lange:2001uf}.
The effect of multiple interactions in the same and neighbouring bunch crossings (pile-up)
was modelled by overlaying~\cite{SIMU-2020-01} the simulated hard-scattering event
with inelastic proton--proton (\(pp\)) events generated from a mix of
\EPOS[2.0.1.4]~\cite{Werner:2005jf} and \PYTHIA[8.308]~\cite{Bierlich:2022pfr}.
The \EPOS events were generated with the \EPOS LHC tune~\cite{Pierog:2013ria}
and the \PYTHIA events with the A3 tune~\cite{ATL-PHYS-PUB-2016-017}
and the \NNPDF[2.3lo]~\cite{Ball:2012cx} set of PDFs.
The MC samples were reweighted so that the pile-up distribution matches the one observed in the data.
All MC samples were processed through an ATLAS detector simulation~\cite{SOFT-2010-01} based on \GEANT~\cite{Agostinelli:2002hh}
or a fast simulation using a parameterisation of the calorimeter response and \GEANT for the other parts of the detector~\cite{SIMU-2018-04}.  All MC samples were reconstructed in the same manner as the data.


%
\section{Event reconstruction}
\label{sec:objects}
%

Events are required to have a reconstructed vertex~\cite{IDTR-2021-01} with at least two associated tracks with $\pt> 500~\MeV$ that are consistent with originating from the beam collision region in the $x$--$y$ plane.  The primary vertex in the event is the vertex with the highest sum of squared transverse momenta of associated tracks.

Jets are reconstructed by applying the \antikt algorithm~\cite{Cacciari:2008gp,Fastjet} with a radius parameter of $R=0.4$ to
topological calorimeter-energy clusters and charged-particle tracks processed with a particle-flow algorithm~\cite{PERF-2015-09}.
The energy of the jets is corrected for the non-compensating calorimeter response, noise threshold effects, energy loss from inactive material, and pile-up contamination~\cite{JETM-2018-05}.
Only jets with $\pT>\SI{20}{\GeV}$ and $|\eta|<4.5$ are considered.
Jet vertex tagger~\cite{PERF-2014-03,PERF-2016-06,ATL-PHYS-PUB-2019-026} algorithms are used to remove jets with $\pT<\SI{60}{\GeV}$ that are identified as not being associated with the event's primary vertex.
Jets with $\pT>\SI{20}{\GeV}$ and $|\eta|<2.5$ containing $b$-hadrons are identified ($b$-tagged) using the GN2 algorithm~\cite{FTAG-2023-05} with a working point (WP) achieving \SI{70}{\percent} efficiency in $\ttbar$ MC events.
To correct for an observed mismodeling of the multiplicity of jets in the forward region in Run 3, simulated events are reweighted as a function of the number of jets in $|\eta| > 2.5$.  The size of this reweighting is approximately 10\% for events with one jet in $|\eta| > 2.5$ and increases to a factor two for events with four jets at high pseudorapidity.
A systematic uncertainty defined as half the effect of this correction is included in the analysis.
Hadronic $\tau$-lepton decays produce a neutrino and visible decay products, mostly one or three charged pions and up to two neutral pions.
The reconstruction of the visible products of hadronically decaying $\tau$-leptons (\tauhadvis) is seeded by jets reconstructed using the \antikt~algorithm where topological clusters of calorimeter energy deposits are used as inputs with a radius parameter of $R=0.4$.  A recursive neural network (RNN) is used to determine if tracks originate from a \tauhadvis, and one (1-prong) or three (3-prong) tracks with a total charge of $\pm 1$ are required~\cite{ATL-PHYS-PUB-2022-044}.

To reject \tauhadvis candidates originating from jets initiated by quarks or gluons, a transformer-based Graph Neural Network (GNN)~\cite{vaswani2023} identification algorithm (GNTau)  is employed.  The architecture is similar to the one described in Ref.~\cite{FTAG-2023-05} and combines track, cluster and jet-level information with global jet features to construct a comprehensive representation of each \tauhadvis candidate.
The model is constructed by concatenating the jet-level feature vector\footnote{A feature vector is a vector whose components are the numerical values of the features describing an event.} with the feature vectors of individual tracks and clusters. The combined inputs are then passed through several dense layers to build an initial latent representation. A transformer layer is subsequently applied to capture contextual dependencies among the constituents, producing context-aware embeddings that encode the complex interactions between all representations. Finally, these refined embeddings are treated as nodes in a fully connected graph, where a graph-based classification step is performed to generate the final output probabilities.
This GNN-based identification algorithm has shown improved discrimination between \tauhadvis and quark/gluon-initiated jets relative to the previous  recurrent neural-network identification algorithm~\cite{ATL-PHYS-PUB-2019-033}, as well as enhanced robustness against pile-up effects.
A `Medium' identification WP,  corresponding to a 75(60)\% efficiency for 1(3)-prong real \tauhad, is used in the analysis.
A separate neural-network discriminant is also constructed to reject backgrounds arising from electrons faking \tauhadvis~\cite{ATL-PHYS-PUB-2022-044}, in this analysis mainly from \ttbar decays with $e$+$\tau$ signatures. This discriminant is built using information from the calorimeter and the tracking detector.

The \tauhadvis energy scale is determined by combining information from the associated tracks, calorimeter energy clusters and reconstructed neutral pions~\cite{PERF-2014-06} using a multivariate regression technique~\cite{ATL-PHYS-PUB-2022-044,ATLAS-CONF-2017-029} trained in MC samples.
The \tauhadvis candidates with  $\pT>20$~\GeV and $|\eta|<2.5$,  excluding the transition region between the endcap and barrel calorimeters  ($1.37<|\eta|<1.52$), are selected.

Electron candidates are reconstructed from energy clusters in the electromagnetic calorimeter matched to a track in the inner detector~\cite{EGAM-2021-01}. They are required to satisfy $\pt>15$~\GeV and $|\eta_{\mathrm{cluster}}|<2.47$, excluding the calorimeter transition region.
The \enquote{\textit{LooseAndBLayer}} electron identification WP~\cite{PERF-2017-01} is used, based on a likelihood discriminant employing calorimeter and tracking variables that provide separation between electrons and jets, and between electrons and photons.
Muon candidates are reconstructed by performing a combined fit of the track information from the inner detector and muon spectrometer~\cite{MUON-2018-03}. They are required to satisfy $\pt > 10~\gev$ and $|\eta| < 2.5$ and the \enquote{\textit{Loose}} identification criteria.

The procedure described in Ref.~\cite{HIGG-2019-09} is employed to remove potential overlaps between the reconstructed jet,  \tauhadvis, electron and muon candidates.
The missing transverse momentum vector, $\vecmet$ (whose magnitude is referred to as missing transverse energy, \met) is reconstructed as the negative vector sum of the transverse momenta of leptons, \tauhadvis and jets, and a soft term, which is calculated as the vectorial sum of the \pT of tracks matched to the primary vertex but not associated with a reconstructed lepton, \tauhadvis or jet~\autocite{JETM-2020-03}.


%
\section{Event selection and categorisation}
\label{sec:evtselection}
\label{sec:evt}

%
%
Events are required to have exactly two \tauhadvis candidates with opposite charge and no electron or muon.
To further reject backgrounds,  events must have $\met>\SI{20}{\GeV}$ and the angular separation between the two \tauhadvis must be $|\Delta\eta|<1.5$ and $0.6<\Delta R<2.5$.
In addition, the two \tauhadvis are required to have the fraction of the $\tau$-lepton's momentum carried by its visible decay products in the collinear approximation~\cite{collinear_approximation,MMCpaper} ($x$) of $0.1<x<1.4$.
To be in the trigger efficiency plateau, the leading (subleading) \tauhadvis is required to have $\pT>40(30)~\SI{}{\GeV}$ and the event must contain at least one jet with $\pT>\SI{70}{\GeV}$ and $|\eta|<3.2$ .
To increase the sensitivity to the \ttH and \tH signals,  events are also required to have at least five jets, with at least one of them being $b$-tagged.
The acceptance times efficiency of this selection is approximately 2\% (1.5\%) for $\ttH$ ($tH$) events with fully hadronic $\ttbar$ and $H\to\tau\tau$ decays.
Background yields are higher in Run~3 compared to Run~2 due to the increased pile-up level in this period.

\begin{table}[bht]
\small
\centering
\caption{Variables used in the multiclass BDT.
The Higgs boson candidate is formed by the vector sum of the two visible $\tau$-lepton candidates' momenta and \vecmet.
In the $p_{\text{T}}$ ratio of jet pairs, the indices denote the $p_{\text{T}}$-ordering of the jets in the event.
}
\renewcommand{\arraystretch}{1.3}
\setlength{\tabcolsep}{10pt}
\scalebox{0.9}{
\begin{tabular}{ll}
\toprule
\textbf{Group} & \textbf{Variable} \\
\midrule

\multirow{5}{*}{\rotatebox{90}{Jet properties}}
& $\eta$ of the 5 leading jets \\
& Scalar sum of all jets $p_{\text{T}}$ \\
& $W$-candidate dijet invariant mass \\
& Top-quark-candidate three-jet invariant mass \\
& Ratio of the $p_{\text{T}}$ of jet pairs ($p_{\text{T}}^{\text{jet,2}} / p_{\text{T}}^{\text{jet,1}}$,  $p_{\text{T}}^{\text{jet,3}} / p_{\text{T}}^{\text{jet,2}}$,  $p_{\text{T}}^{\text{jet,4}} / p_{\text{T}}^{\text{jet,2}}$) \\
\midrule

\multirow{4}{*}{\rotatebox{90}{$\Delta R$, $\Delta\eta$}}
& Minimum $\Delta R$ between two jets \\
& Minimum $\Delta R$ between a $b$-tagged jet and a $\tau$-lepton \\
& $|\Delta\eta(\tau,\tau)|$ \\
& $\Delta R(\tau,\tau)$ \\
\midrule

\multirow{4}{*}{\rotatebox{90}{$\tau$-lepton}}
& $p_{\text{T}}(\tau\tau)$ \\
& Sub-leading $\tau$ $p_{\text{T}}$ \\
& Leading $\tau$ $\eta$ \\
& Di-$\tau$ invariant mass (\mmc) \\
\midrule

\multirow{2}{*}{\rotatebox{90}{\etmiss}}
& Missing transverse energy $E_{T}^{\text{miss}}$ \\
& Smallest $\Delta\phi(\tau, \vec{E}_{T}^{\text{miss}})$ \\
\midrule

\multirow{8}{*}{\rotatebox{90}{$t\bar{t}H$ vs.\ $tH$}}
& Number of light (non-$b$-tagged) jets \\
& Number of light jets with the same $\eta$ sign \\
& $b$-jets multiplicity \\
& $\eta$ of the leading $b$-jet \\
& $p_{\text{T}}$ of the leading $b$-jet  \\
& Largest $\Delta \eta$ between any combination of light-jet and $b$-jet  \\
& Maximum $\Delta \eta$ between the visible Higgs boson decay products and any $b$-jet \\
& Largest invariant mass of any pair of light jets \\
\bottomrule
\end{tabular}
}
\label{tab:all_vars}
\end{table}

To enhance the signal to background separation,  a multiclass Boosted Decision Tree (BDT) trained with TMVA~\cite{TMVA} is used.
This BDT is trained using four separate classes: two for the signals (\ttH and $tH$) and two for the main backgrounds (\Ztt and \ttbar).
Table~\ref{tab:all_vars} shows all the BDT input variables.  Some of these variables were already used in Ref.~\cite{HIGG-2022-07},  such as the properties of $\tau$-leptons,  jets and $\met$,  the mass of the $W$-boson candidate built as the pair of non-$b$-tagged jets in the event with invariant mass closest to $m_W$,  or the  mass of the top-quark candidate built as the system of the $W$-boson candidate and a $b$-tagged jet with invariant mass closest to \mtop.
Several extra variables are added to increase the sensitivity for $tH$ as shown in Table~\ref{tab:all_vars}.
Furthermore, the di-$\tau$ invariant mass ($m_{\tau\tau}$) computed with the Missing Mass Calculator (MMC) technique~\cite{MMCpaper} is also used as input in the BDT training.
The modelling of the MC simulation for the BDT input variables and output scores is checked in data after the selection described above, as well as in regions with $m_{\tau\tau}<\SI{100}{\GeV}$ (enhanced in \Ztt) and with $m_{\tau\tau}>\SI{150}{\GeV}$  (enhanced in \ttbar).

Events are classified in four different categories (\ttH, $tH$, \Ztt or \ttbar) according to their highest BDT score among the four possible classes.
The \ttH and $tH$ signal regions are also required to have $100 <m_{\tau\tau} < \SI{150}{\GeV}$, while events outside this window are assigned to
sideband regions that are used to further control the backgrounds.
For the \ttH signal regions, a further split into three STXS \pTH categories, $\pTH \in [0, 200, 300, \infty)$, is performed, with \pTH estimated with the neural network described in Ref.~\cite{HIGG-2022-07}.
The BDT output scores are used as discriminants in the statistical analysis.
Contributions in the signal regions from other Higgs boson decay modes besides \Htautau are negligible.

%

%


%
\section{Background estimation}
\label{sec:bkg}


The main backgrounds affecting this analysis arise from $\Ztt$ and \ttbar production, as well as the reducible background from processes where at least one jet is misidentified as a \tauhadvis (collectively referred to as misidentified $\tau$ background).

The contribution from the misidentified $\tau$ background is estimated with a fake-factor method~\cite{EGAM-2019-01,HIGG-2019-09}.
Data events are selected if they satisfy a very loose requirement on the \tauhadvis identification score but do not satisfy the \enquote{Medium} WP requirements.
Residual contributions from processes with real \tauhadvis satisfying this requirement are evaluated using simulations and subtracted.
The distribution of the misidentified \tauhad background is obtained by multiplying the contribution of data events selected with the inverted identification criterion by a fake-factor defined as the ratio of misidentified \tauhadvis that respectively satisfy or not the \enquote{Medium} identification WP.
The fake-factors are measured in control regions enhanced in $W$+jets production that require the presence of one electron or muon plus one \tauhadvis candidate.
These fake-factors are measured separately for Run~2 and Run 3 in bins of ($\pT$, $\eta$) and their values are in the range of 0.05--0.35.
The correctness of the background estimate is assessed in validation regions requiring two \tauhadvis with the same electric charge, where the estimated background agrees with the observed data within 10\%.

Dedicated control regions are built for the irreducible $\Ztt$ and \ttbar backgrounds using the BDT scores described in Section~\ref{sec:evt}.
The prediction from MC for these processes is normalised using the data observation within the likelihood fit described in Section~\ref{sec:result}.  Separate control regions and normalisation factors are used for Run~2 and Run 3.
Figure~\ref{fig:postfit_crs} shows background predictions in each bin of the control regions after the fit described in Section~\ref{sec:result},  achieving a good agreement with the observed data.

\begin{figure}[!phtb]
\centering
\subfloat[]{\includegraphics[width=0.85\linewidth]{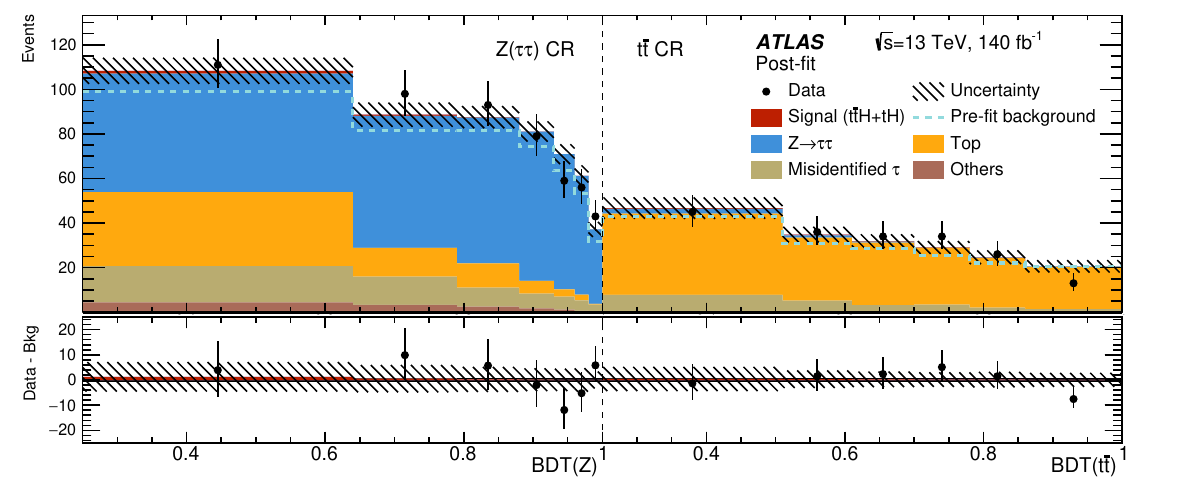}}\\
\subfloat[]{\includegraphics[width=0.85\linewidth]{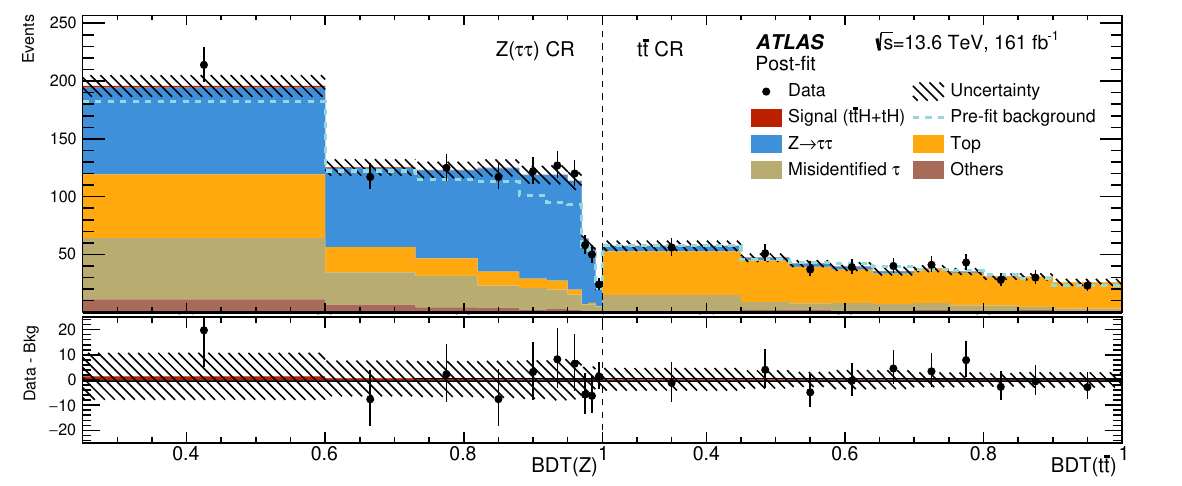}}
\caption{
Comparison of the observed and expected event yields in the \Ztt and \ttbar control regions as a function of their corresponding BDT score,  showing all bins used in the statistical analysis for (a) Run 2 and (b) Run 3.
The signal and background normalisation factors and the nuisance parameters are set to their best-fit values, while the pre-fit background prediction is shown as a dashed line.
The signal distribution includes the contributions from \ttH and \tH.
The shaded band indicates the total uncertainty, including correlations.
The bottom panel shows the differences between observed data events and expected background.
}
\label{fig:postfit_crs}
\end{figure}


%

%
\section{Systematic uncertainties}
\label{sec:syst}
%


Systematic uncertainties are considered for the experimental procedures,  the signal and
background cross-sections, and the modelling of the relevant physics processes.
In each region,  uncertainties can affect the normalisations, the shape of the BDT discriminants used in the fit, or both.
Correlations of the systematic uncertainties in signal and background are taken into account.

The primary sources of systematic uncertainties are related to the jet energy calibration and resolution~\cite{JETM-2018-05},
the $\ttbar$ modelling uncertainties,  uncertainties on the $\Ztt$ and \ttbar normalisation factors, and on the misidentified $\tau$ background.
Uncertainties in the $t\bar{t}$ parton shower are assessed by using a \POWHER[7]~\cite{Bahr:2008pv,Bellm:2015jjp} $t\bar{t}$ MC simulation as a variation of the nominal \POWPY[8] prediction.
Uncertainties related to the \ttbar MC sample normalisation and factorisation scales,  initial- and final-state radiation,  matching between the matrix element and parton shower~\cite{Hoche:2021mkv,ATL-PHYS-PUB-2023-029} and \hdamp\ parameters are also included.
For $\ttH$ and $tH$ production, systematic uncertainties on PDF,  $\alphas$ and QCD scales are considered.
Additional uncertainties for $\ttH$ matrix element and parton shower are evaluated with alternative generators.
Conservative uncertainties of 100\% are applied to the contributions from other Higgs boson production modes arising from imprecise modelling of associated heavy-flavour jet production.

The systematic uncertainties in the estimation of the misidentified $\tau$ background closely follow the strategy in Ref.~\cite{HIGG-2019-09},  including the statistical uncertainties in the fake-factors,  uncertainties due to their parametrisation,  and uncertainties related to the variation between fake-factors calculated in different regions.

A minor impact in the analysis is found from the experimental uncertainties on \tauhadvis and $b$-tagging.  For the $\tauhadvis$ identification,  energy calibration and trigger efficiency,  uncertainties derived on the previous generation identification algorithm~\cite{ATL-PHYS-PUB-2019-033} are considered.

All systematic uncertainties are treated as uncorrelated between Run 2 and Run 3 except for a subset of
uncertainties associated with jet energy calibration, and background and signal theoretical systematic uncertainties.

%


%
\section{Results}
\label{sec:result}
%


A combined profile-likelihood fit is performed on all bins in the signal and control regions to simultaneously measure the $\ttH$ and $tH$ cross-sections  and the normalisation of the different background processes.
Similar fits are also performed to measure the $\ttH$ cross-section in STXS bins.
The statistical analysis is performed using \AtlasMC{TRExFitter}~\cite{TRExFitter} with a likelihood function constructed as the product of Poisson probability terms ($P$) over the bins of the input distributions:

\begin{equation}
\label{eqn:llh}
L\left(\vec{\mu},  \vec{\lambda}, \vec{\theta}, \text{data}\right) = \prod_{k\in\text{reg}}
\prod_{b\in\text{bins}}   P(n_{k,b} | n^{\text{signal}}_{k,b}(\vec{\mu}, \vec{\theta}) +
n^{\text{bkg}}_{k,b}(\vec{\lambda}, \vec{\theta})) \prod_{\theta\in\vec{\theta}} G(\theta),
\end{equation}

where $n_{k,b}$, $n^{\text{signal}}_{k,b}$ and $n^{\text{bkg}}_{k,b}$
stand for the number of observed events, the expected number of signal events and the expected number of background events in bin $b$ of analysis region $k$, respectively.
The systematic uncertainties (described in Section~\ref{sec:syst}) are incorporated in the likelihood function through nuisance parameters, $\vec{\theta}$, which are constrained by Gaussian probability terms, $G$.
The statistical uncertainty due to the limited number of simulated events is included in the likelihood in the form of additional nuisance parameters with Poisson constraint terms.
Both $\vec{\mu}$ (signal-strength parameter,  defined as a multiplicative factor to the yield of expected signal events) and $\vec{\lambda}$ (normalisation factors for several backgrounds) are treated as free parameters in the likelihood fit and are estimated by maximising the likelihood.
The test statistic is constructed from the profile likelihood ratio, and the confidence intervals for the parameters of interest are
derived using the asymptotic approximation~\cite{Cowan:2010js}.

\begin{figure}[ptbh]
\centering
\subfloat[]{\includegraphics[width=0.95\linewidth]{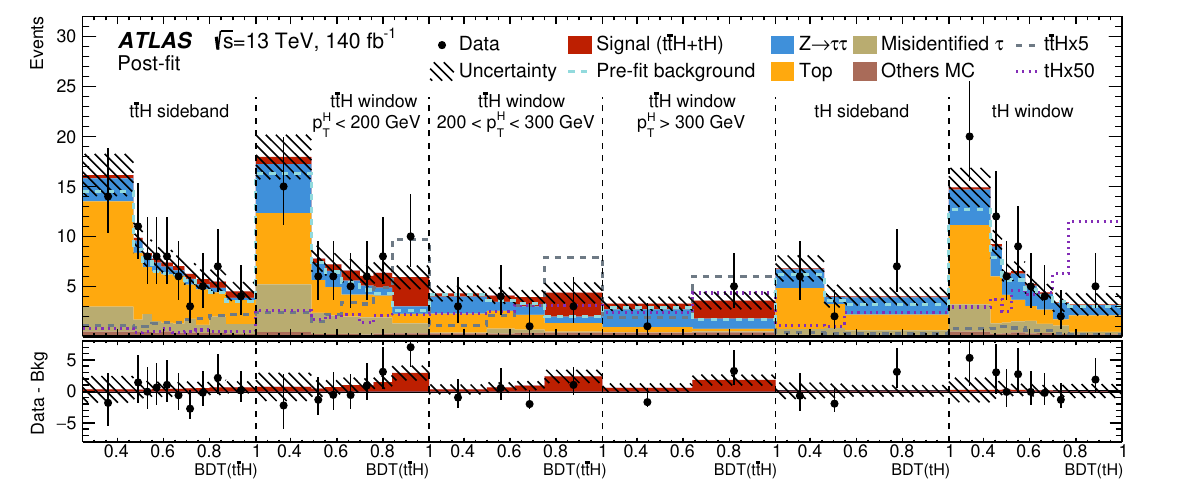}}\\
\subfloat[]{\includegraphics[width=0.95\linewidth]{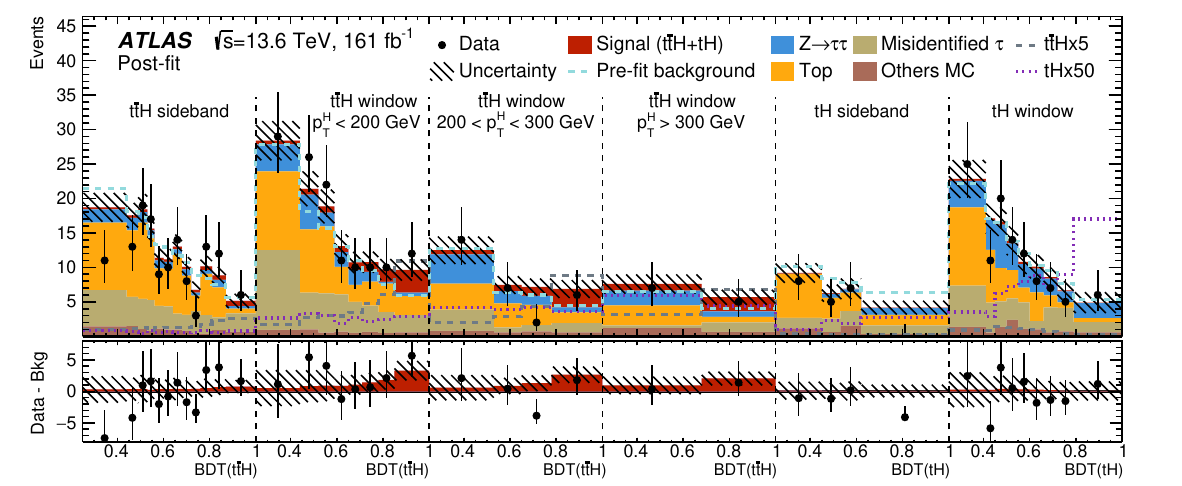}}
\caption{
Comparison of the observed and expected event yields in all the signal regions and bins used in the statistical analysis for (a) Run 2 and (b) Run 3.
The regions labelled as `window' correspond to a $100<\mmc<\SI{150}{\GeV}$ while those labelled as `sideband' correspond to either $\mmc<\SI{100}{\GeV}$ or $\mmc>\SI{150}{\GeV}$.
The signal and background normalisation factors and the nuisance parameters are set to their best-fit values, while the pre-fit background prediction is shown as a dashed line.
The signal distribution includes the contributions from \ttH and \tH.
The shaded band indicates the total uncertainty, including correlations.
The bottom panel shows the differences between observed data events and expected background.
Dashed lines illustrating the SM expected \ttH and \tH contributions scaled by a factor 5 and 50, respectively, are also shown.
}
\label{fig:postfit_srs}
\end{figure}

Figure~\ref{fig:postfit_srs} shows the data, estimated background and observed signal in each bin of the signal regions.
The background normalisation factors in the fit are consistent with unity and similar between Run~2 and Run~3.
The measurements of the signal strength for $\ttH$ and $tH$ production with \Htt are shown in Figure~\ref{fig:pois}.
The best-fit values of the signal strengths are $\mu_{\ttH}=1.51^{+0.71}_{-0.62}$ and $\mu_{tH}=-0.4^{+5.6}_{-5.1}$
with a $-34\%$ correlation between them.
The compatibility of this fit with the SM is 84\%.
The observed (expected) significance for $\ttH$ production is 2.6 (1.8) standard deviations.
The measured $\mu_{tH}$ value is consistent with previous ATLAS results~\cite{HIGP-2026-05,HIGP-2024-03,HIGP-2024-08}.
Figure~\ref{fig:contours} shows the results as a function of the $\ttH$ and $tH$ signal strength parameters.

Figure~\ref{fig:stxs} shows the measured $\ttH$ cross-section in the  STXS  framework.  In this fit,  $tH$ is also allowed to float freely.  Good agreement with the SM prediction is observed within the uncertainties of the measurement and the compatibility of the STXS fit with the SM is 52\%.

\begin{figure}[!phtb]
\centering
\includegraphics[width=0.6\linewidth]{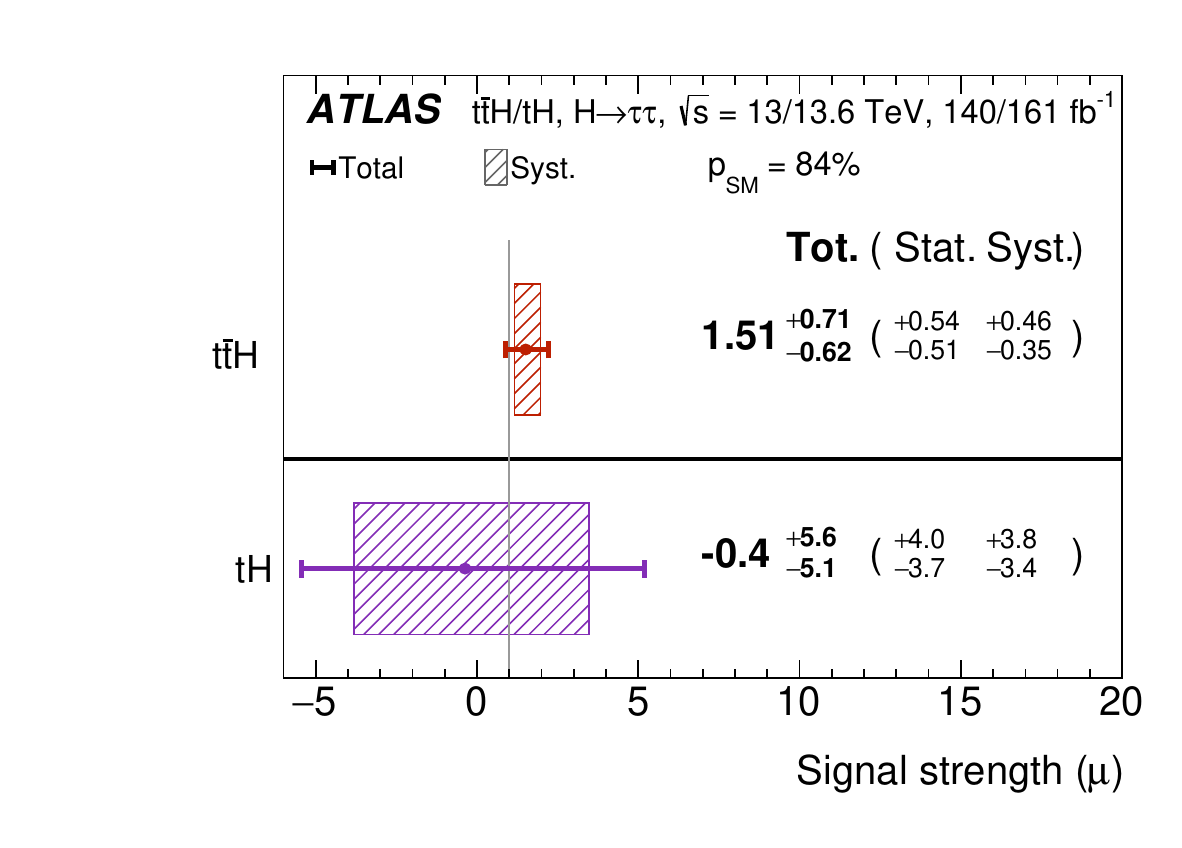}
\caption{
Measured values of the signal strength for $\ttH$ and $tH$ production.  The $\pm 1\sigma$ uncertainty in each measurement is indicated by the error bar, with the contribution from the systematic uncertainty indicated by the coloured band.
}
\label{fig:pois}
\end{figure}

\begin{figure}[!phtb]
\centering
\includegraphics[width=0.5\linewidth]{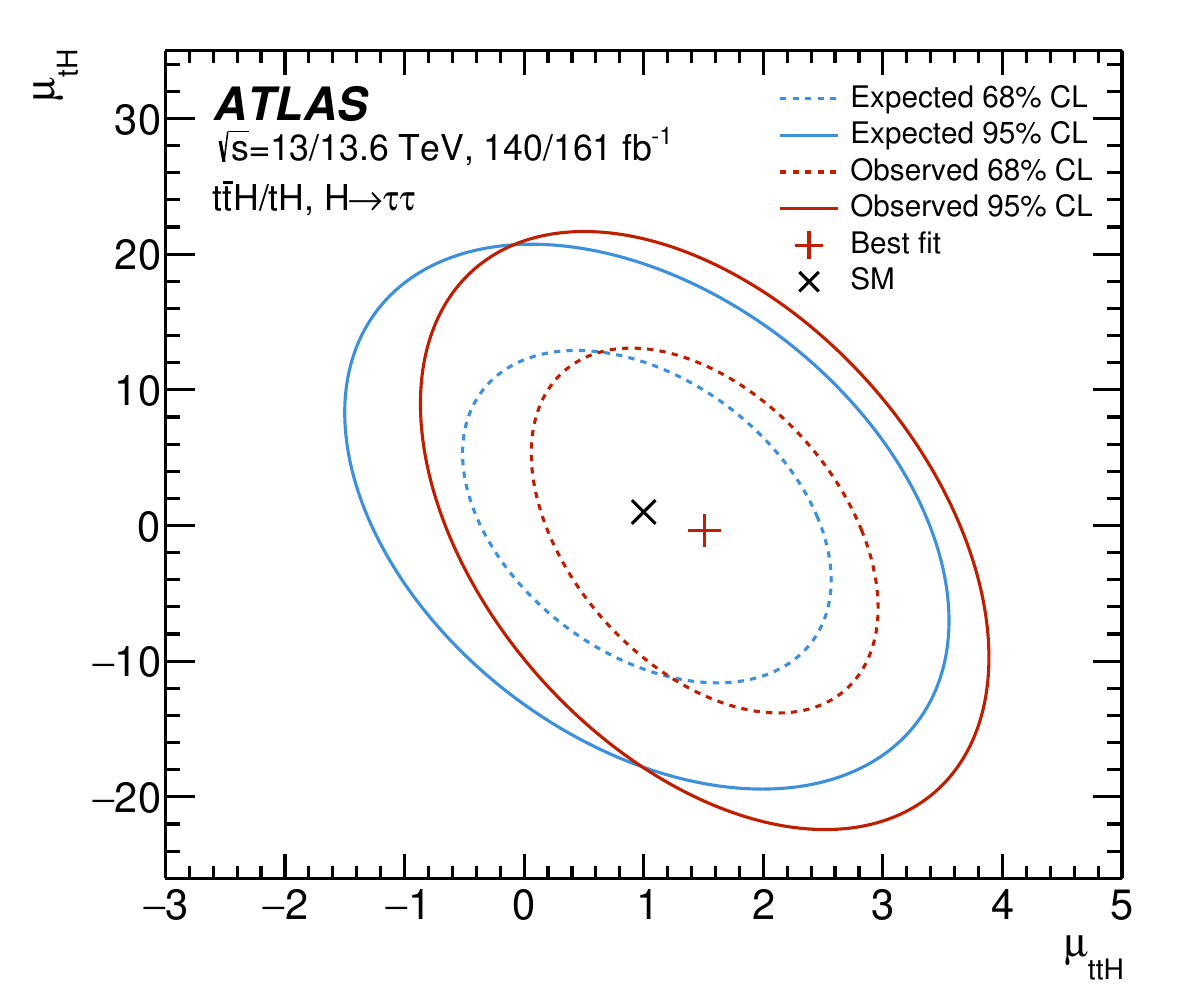}
\caption{
Observed and expected best-fit $\mu_{\ttH}$ and $\mu_{tH}$ signal strengths, with dashed (solid) lines showing their 68\% (95\%) confidence level (CL) contours.
}
\label{fig:contours}
\end{figure}

\begin{figure}[!phtb]
\centering
\includegraphics[width=0.65\linewidth]{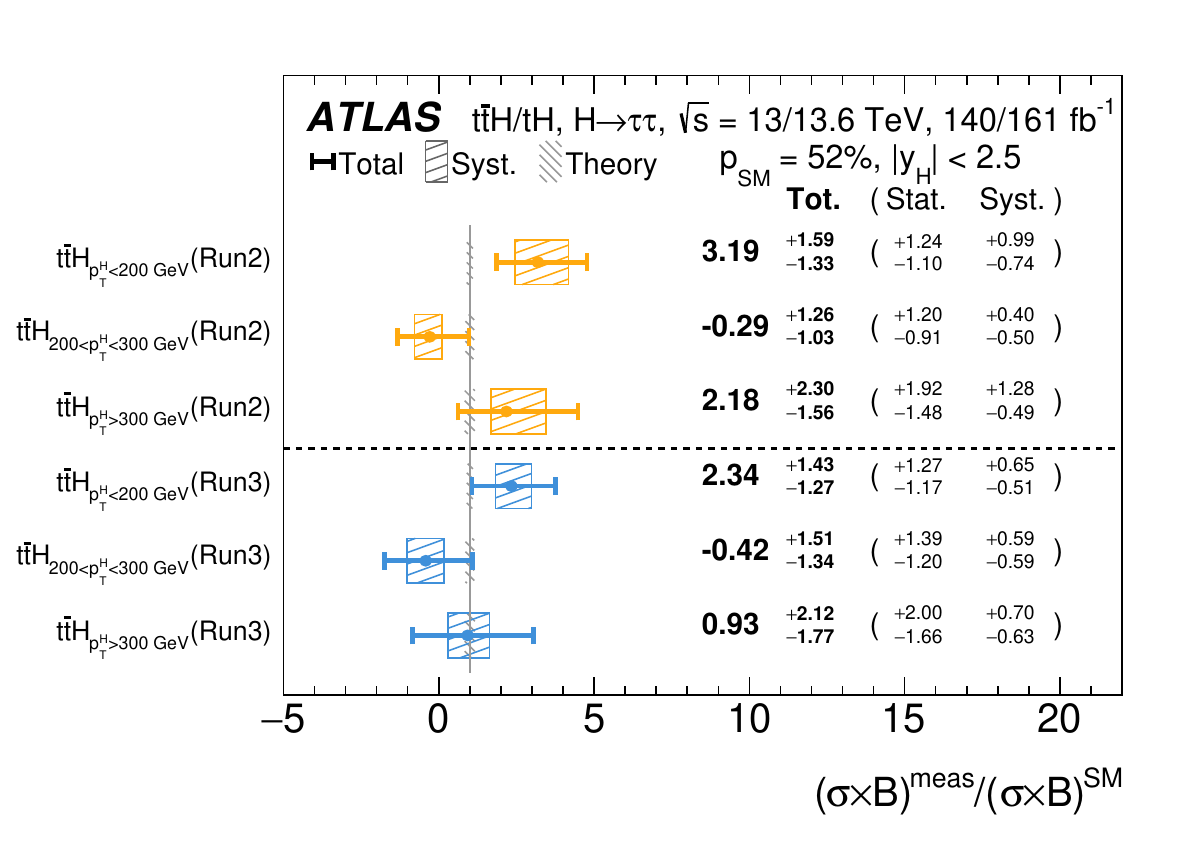}
\caption{
Measured values of $\sigma \times B(H\to\tau\tau)$ relative to the SM expectations in the $\ttH$ simplified template cross-section bins of $\pTH$.  The $\pm 1\sigma$ uncertainty in each measurement is indicated by the error bar, with the contribution from the systematic uncertainty indicated by the coloured band.
Uncertainties of the SM prediction are also shown.
}
\label{fig:stxs}
\end{figure}


%
%
\FloatBarrier

\section{Conclusion}
\label{sec:conclusion}
%

A measurement of the Higgs boson production cross-section in association with either a top-quark pair (\ttH) or a single top quark ($tH$) in the $\Htt$ decay channel has been presented.
Only final states with fully hadronic decays of top quark and $\tau$-leptons are considered.
The analysis uses proton--proton collisions at $\sqrt{s}=13$ and $\SI{13.6}{\TeV}$ corresponding to integrated luminosities of 140~fb$^{-1}$ and 161~fb$^{-1}$, respectively,  recorded by the ATLAS detector at the LHC.
Results benefit from the newly developed GNTau $\tau$-lepton identification algorithm, which is used for the first time in ATLAS,  as well as from a more refined event classification and the increase in data statistics.
The signal strength ($\mu$), defined as the observed cross-section normalised to SM prediction, is measured to be
$\mu_{t\bar{t}H}=1.51^{+0.71}_{-0.62}$ for \ttH and $\mu_{tH}=-0.4^{+5.6}_{-5.1}$ for $tH$.
The \ttH production cross-section is also measured differentially in bins of the Higgs boson transverse momentum within the
simplified template cross-section framework.
In all cases,  results are consistent with existing measurements and with the SM predictions.


%
%
%
%
%
\section*{Acknowledgements}
%

%

%
%

%
%

We thank CERN for the very successful operation of the LHC and its injectors, as well as the support staff at
CERN and at our institutions worldwide without whom ATLAS could not be operated efficiently.

The crucial computing support from all WLCG partners is acknowledged gratefully, in particular from CERN, the ATLAS Tier-1 facilities at TRIUMF/SFU (Canada), NDGF (Denmark, Norway, Sweden), CC-IN2P3 (France), KIT/GridKA (Germany), INFN-CNAF (Italy), NL-T1 (Netherlands), PIC (Spain), RAL (UK) and BNL (USA), the Tier-2 facilities worldwide and large non-WLCG resource providers. Major contributors of computing resources are listed in Ref.~\cite{ATL-SOFT-PUB-2026-001}.

We gratefully acknowledge the support of ANPCyT, Argentina; YerPhI, Armenia; ARC, Australia; BMWFW and FWF, Austria; ANAS, Azerbaijan; CNPq and FAPESP, Brazil; NSERC, NRC and CFI, Canada; CERN; ANID, Chile; CAS, MOST, NSFC and FRFCU, China; Minciencias, Colombia; MEYS CR, Czech Republic; DNRF and DNSRC, Denmark; IN2P3-CNRS and CEA-DRF/IRFU, France; SRNSFG, Georgia; BMFTR, HGF and MPG, Germany; GSRI, Greece; RGC and Hong Kong SAR, China; ICHEP and Academy of Sciences and Humanities, Israel; INFN, Italy; MEXT and JSPS, Japan; CNRST, Morocco; NWO, Netherlands; RCN, Norway; MNiSW, Poland; FCT, Portugal; MNE/IFA, Romania; MSTDI, Serbia; MSSR, Slovakia; ARIS and MVZI, Slovenia; DSI/NRF, South Africa; MICIU/AEI, Spain; SRC and Wallenberg Foundation, Sweden; SERI, SNSF and Cantons of Bern and Geneva, Switzerland; NSTC, Taipei; TENMAK, T\"urkiye; STFC/UKRI, United Kingdom; DOE and NSF, United States of America.

Individual groups and members have received support from BCKDF, CANARIE, CRC and DRAC, Canada; CERN-CZ and PRIMUS, Czech Republic; COST, ERC, ERDF, Horizon 2020 and Marie Sk{\l}odowska-Curie Actions, European Union; Investissements d'Avenir Labex, Investissements d'Avenir Idex and ANR, France; DFG and AvH Foundation, Germany; Herakleitos, Thales and Aristeia programmes co-financed by EU-ESF and the Greek NSRF, Greece; BSF-NSF and MINERVA, Israel; NCN and NAWA, Poland; La Caixa Banking Foundation, CERCA and AGAUR programs from Generalitat de Catalunya and PROMETEO and GenT Programmes Generalitat Valenciana, Spain; G\"{o}ran Gustafssons Stiftelse, Sweden; The Royal Society and Leverhulme Trust, United Kingdom; Eric and Wendy Schmidt Fund for Strategic Innovation, United States of America.

In addition, individual members wish to acknowledge support from Chile: Agencia Nacional de Investigaci\'on y Desarrollo (ANID FONDECYT reg. 1230987, FONDECYT 1230812, FONDECYT 1240864, Fondecyt 3240661, Fondecyt Regular 1240721); China: Fundamental Research Funds for the Central Universities (010-63263105), Chinese Ministry of Science and Technology (MOST-2023YFA1605700, MOST-2023YFA1609300), National Natural Science Foundation of China (NSFC 12275265, NSFC-W2543005); Czech Republic: Czech Science Foundation (GACR - 24-11373S), Ministry of Education Youth and Sports (ERC-CZ-LL2327, FORTE CZ.02.01.01/00/22\_008/0004632); EU: H2020 European Research Council (ERC - 101002463); European Union: European Research Council (BARD No. 101116429, ERC - 101219398, ERC 101089007), European Regional Development Fund (HE COFUND GA No.101081355, ERDF), Marie Sklodowska-Curie Actions (GAP-101168829); France: Agence Nationale de la Recherche (ANR-22-EDIR-0002, ANR-24-CE31-0504-01); Germany: Deutsche Forschungsgemeinschaft (DFG - 469666862); China: Research Grants Council (GRF); Italy: Istituto Nazionale di Fisica Nucleare (LHC-MIUR - 28003/2025), Ministero dell'Università e della Ricerca (NextGenEU I53D23000820006 M4C2.1.1, SOE2024\_0000023); Japan: Japan Society for the Promotion of Science (JSPS KAKENHI  JP25H0063, JSPS KAKENHI JP22H04944, JSPS KAKENHI JP24K23939, JSPS KAKENHI JP24KK0251, JSPS KAKENHI JP25H00650, JSPS KAKENHI JP25H01291, JSPS KAKENHI JP25K01011, JSPS KAKENHI JP25K01023, JSPS KAKENHI JP25KK0047); Poland: Polish National Science Centre (NCN 2021/42/E/ST2/00350, NCN OPUS 2023/51/B/ST2/02507, NCN OPUS nr 2022/47/B/ST2/03059, NCN UMO-2019/34/E/ST2/00393, UMO-2022/47/O/ST2/00148, UMO-2023/49/B/ST2/04085, UMO-2023/51/B/ST2/00920, UMO-2024/53/N/ST2/00869); Spain: Agència de Gestió d'Ajuts Universitaris i de Recerca. (AGAUR - 2023 BP 00141), Ministry of Science and Innovation (RYC2019-028510-I, RYC2020-030254-I, RYC2021-031273-I, RYC2022-038164-I), Ministerio de Ciencia, Innovación y Universidades/Agencia Estatal de Investigaci\'on (EU NextGenerationEU (PRTR-C17.I1), PID2022-142604OB-C22); Sweden: Carl Trygger Foundation (Carl Trygger Foundation CTS 22:2312), Swedish Research Council (Swedish Research Council 2023-04654, VR 2021-03651, VR 2022-03845, VR 2022-04683, VR 2023-03403, VR 2024-05451, VR 2025-05940), Knut and Alice Wallenberg Foundation (KAW 2023.0366); United Kingdom: The Binks Trust, Royal Society (NIF-R1-231091); United States of America: U.S. Department of Energy (ECA DE-AC02-76SF00515), John Templeton Foundation (John Templeton Foundation 63206), Neubauer Family Foundation.

%
%


%
%
\printbibliography
\clearpage
\input{atlas_authlist}
%

%
%

%

%
%
%
%

\end{document}

%% file: atlas_authlist.tex
 
\begin{flushleft}
\hypersetup{urlcolor=black}
{\Large The ATLAS Collaboration}

\bigskip

\AtlasOrcid[0000-0002-6665-4934]{G.~Aad}$^\textrm{\scriptsize 102}$,
\AtlasOrcid[0000-0001-7616-1554]{E.~Aakvaag}$^\textrm{\scriptsize 17}$,
\AtlasOrcid[0000-0002-5888-2734]{B.~Abbott}$^\textrm{\scriptsize 121}$,
\AtlasOrcid[0000-0002-0287-5869]{S.~Abdelhameed}$^\textrm{\scriptsize 83b}$,
\AtlasOrcid[0000-0002-1002-1652]{K.~Abeling}$^\textrm{\scriptsize 54}$,
\AtlasOrcid[0000-0001-5763-2760]{N.J.~Abicht}$^\textrm{\scriptsize 48}$,
\AtlasOrcid[0000-0002-8496-9294]{S.H.~Abidi}$^\textrm{\scriptsize 30}$,
\AtlasOrcid[0009-0003-6578-220X]{M.~Aboelela}$^\textrm{\scriptsize 44}$,
\AtlasOrcid[0000-0002-9987-2292]{A.~Aboulhorma}$^\textrm{\scriptsize 36e}$,
\AtlasOrcid[0000-0001-5329-6640]{H.~Abramowicz}$^\textrm{\scriptsize 154}$
% ...
\bigskip
\\

% ...
$^{17}$Department for Physics and Technology, University of Bergen, Bergen; Norway.\\
% ...
$^{30}$Physics Department, Brookhaven National Laboratory, Upton NY; United States of America.\\
% ...
$^{36}$$^{(a)}$Facult\'e des Sciences Ain Chock, Universit\'e Hassan II de Casablanca;$^{(b)}$Facult\'{e} des Sciences, Universit\'{e} Ibn-Tofail, K\'{e}nitra;$^{(c)}$Facult\'e des Sciences Semlalia, Universit\'e Cadi Ayyad, LPHEA-Marrakech;$^{(d)}$LPMR, Facult\'e des Sciences, Universit\'e Mohamed Premier, Oujda;$^{(e)}$Facult\'e des sciences, Universit\'e Mohammed V, Rabat;$^{(f)}$Institute of Applied Physics, Mohammed VI Polytechnic University, Ben Guerir; Morocco.\\
% ...
$^{44}$Physics Department, Southern Methodist University, Dallas TX; United States of America.\\
% ...
$^{48}$Fakult\"{a}t Physik, Technische Universit{\"a}t Dortmund, Dortmund; Germany.\\
% ...
$^{54}$II. Physikalisches Institut, Georg-August-Universit\"{a}t G\"ottingen, G\"ottingen; Germany.\\
% ...
$^{83}$$^{(a)}$Khalifa University of Science and Technology, Abu Dhabi;$^{(b)}$New York University Abu Dhabi, Abu Dhabi;$^{(c)}$United Arab Emirates University, Al Ain;$^{(d)}$University of Sharjah, Sharjah; United Arab Emirates.\\
% ...
$^{102}$CPPM, Aix-Marseille Universit\'e, CNRS/IN2P3, Marseille; France.\\
% ...
$^{121}$Homer L. Dodge Department of Physics and Astronomy, University of Oklahoma, Norman OK; United States of America.\\
% ...
$^{154}$Raymond and Beverly Sackler School of Physics and Astronomy, Tel Aviv University, Tel Aviv; Israel.\\
% ...
$^{a}$ Also at Affiliated with an institute formerly covered by a cooperation agreement with CERN.\\
$^{b}$ Also at An-Najah National University, Nablus; Palestine.\\
$^{c}$ Also at Borough of Manhattan Community College, City University of New York, New York NY; United States of America.\\
$^{d}$ Also at Center for Interdisciplinary Research and Innovation (CIRI-AUTH), Thessaloniki; Greece.\\
$^{e}$ Also at Centre of Physics of the Universities of Minho and Porto (CF-UM-UP); Portugal.\\
$^{f}$ Also at CERN, Geneva; Switzerland.\\
% ...

\end{flushleft}
